\documentclass[twocolumn]{autart}    

\usepackage{graphicx}          

  \ifx\pdfoutput\undefined 
\usepackage{graphicx}
\DeclareGraphicsExtensions{.eps,.ps} \else
\usepackage{graphicx}
\DeclareGraphicsExtensions{.pdf} \fi
\usepackage{amssymb}
\newtheorem{theorem}{Theorem}
\newtheorem{lemma}{Lemma}
\newtheorem{remark}{Remark}

\usepackage{color,xcolor}
\usepackage{fancyhdr}
\begin{document}

\begin{frontmatter}

\title{Robust Distributed Maximum Likelihood Estimation with Dependent Quantized Data} 

\thanks{This paper was not presented at any IFAC
meeting. Corresponding author P. K. Varshney Tel. (315) 443-1060.
Fax (315) 443-4745. Xiaojing Shen is with Department of Mathematics,
Sichuan University, Chengdu, Sichuan 610064, China. He was with
Syracuse University during 2012--2013. This work was supported  in
part by  U.S. Air Force Office of Scientific Research (AFOSR) under
Grants FA9550-10-1-0263 and FA9550-10-1-0458 and in part by the NNSF
of China 61273074 and  IRT1273.}

\author[Chengdu]{Xiaojing Shen}\ead{shenxj@scu.edu.cn},    
\author[Syracuse]{Pramod K. Varshney}\ead{varshney@syr.edu},               
\author[Chengdu]{Yunmin Zhu}\ead{ymzhu@scu.edu.cn}  
\address[Chengdu]{Department of Mathematics,
Sichuan University, Chengdu, Sichuan 610064, China.}        
\address[Syracuse]{ Department of Electrical
Engineering and Computer Science, Syracuse University, NY, 13244,
USA. }  

\begin{keyword}                           
Maximum likelihood estimation; distributed
estimation; Fisher information matrix; wireless sensor networks             
\end{keyword}                             

\begin{abstract}                          
In this paper, we consider distributed maximum likelihood estimation
(MLE) with dependent quantized data under the assumption that the
structure of the joint probability density function (pdf) is known,
but it contains unknown deterministic parameters. The parameters may
include different vector  parameters corresponding to marginal pdfs
and  parameters that describe dependence of observations across
sensors. Since MLE with a single quantizer is sensitive to the
choice of thresholds due to the uncertainty of pdf, we concentrate
on MLE with multiple groups of quantizers (which can be determined
by the use of prior information or some heuristic approaches) to
fend off against the risk of a poor/outlier quantizer. The
asymptotic efficiency of the MLE scheme with multiple quantizers is
proved under some regularity conditions and the asymptotic variance
is derived to be the inverse of a weighted linear combination of
Fisher information matrices based on multiple different quantizers
which can be used to show the robustness of our approach. As an
illustrative example, we consider an estimation problem with a
bivariate non-Gaussian pdf that has applications in distributed
constant false alarm rate (CFAR) detection systems. Simulations show
the robustness of the proposed MLE scheme especially when the number
of quantized measurements is small.
\end{abstract}

\end{frontmatter}

\section{Introduction}
Wireless sensor networks  have attracted much attention with a lot
of research taking place over the past several years. Many
 advances have been made in distributed detection,
estimation, tracking and control (see e.g.,
\cite{Veeravalli-Varshney12} and references therein).
Distributed estimation and quantization problems have been
considered in a number of previous studies. The parameters to be
estimated are modeled as \emph{random} and \emph{deterministic }in
different situations. For \emph{random} parameters, there exist
various prior studies under the assumption of known joint pdf of
parameters and sensor measurements (see, e.g.,
\cite{Megalooikonomou-Yesha00}).
We concentrate on  \emph{deterministic }parameters in this paper.
For deterministic parameters, several universal distributed
estimation schemes have been proposed
\cite{Xiao-Ribeiro-Luo-Giannakis06} in the presence of unknown,
additive sensor noises that are bounded and identically distributed.
The work in \cite{Ribeiro-Giannakis05} addressed design
and implementation issues under the assumption of a \emph{scalar}
parameter to be estimated and using \emph{scalar} quantizers.
The work in \cite{Fang-Li09} proposed \emph{vector} quantization
design for distributed estimation under the assumption of additive
observation noise model.

System identification  based on quantized measurements is a
challenging problem even for very simple models and has been
researched for a wide range of applications (see, e.g.,
\cite{Wang-Yin-Zhang-Zhao10}). A method for recursive identification
of the nonlinear Wiener model was developed in \cite{Wigren95} and
the corresponding convergence properties were analyzed. In
\cite{Godoy-Goodwin-Aguero-Marelli-Wigren11}, Godoy et al. developed
an MLE approach and used a scenario-based form of the expectation
maximization algorithm to parameter estimation for general MIMO FIR
linear systems with quantized outputs.  The problem of set
membership system identification with quantized measurements was
considered in \cite{Casini-Garulli-Vicino12}. In
\cite{Gustafsson-Karlsson09}, the results from statistical
quantization theory were surveyed and applied to both moment
calculations and the likelihood function of the measured signal. The
system identification of ARMA models using intermittent and
quantized output observations was proposed in
\cite{Marelli-You-Fu13}. The formal conditions for the asymptotic
normality of the  MLE  to the reliability of a complex system based
on a combination of full system and subsystem tests were proposed in
\cite{Spall12}.

In previous works, the MLE with quantized data is extensively used
to estimate the deterministic parameters.
In this paper, robust
distributed MLE with dependent quantized data is considered. Our
work differs from previous studies in several aspects. Previous
results concentrate on the problem of how to design the quantization
schemes for estimating a deterministic parameter where each sensor
makes \emph{one} noisy observation. The observations are usually
assumed \emph{independent} across sensors, and they discuss the
relationship between MLE performance and the number of sensors.
Here, we focus on the problem of how to design estimation schemes
for  the unknown parameter \emph{vector} associated with the joint
pdf of the observations where the number of sensors is \emph{fixed}.
The emphasis here is on system robustness. These observations may be
\emph{dependent} across sensors. The unknown parameters may include
different vector parameters corresponding to marginal pdfs and
parameters that describe dependence of observations across sensors.
Actually, the dependence between sensors is very important in
multisensor fusion systems, for example, see the recent work on
distributed location estimation with dependent sensor observations
\cite{Sundaresan-Varshney11}.


In this paper, we investigate the performance of MLE with multiple
quantizers, since MLE with a single quantizer is sensitive to the
choice of thresholds due to the uncertainty of pdf (see, e.g.,
\cite{Fang-Li08}). Our main contribution is that we analytically
derive the \emph{asymptotic efficiency}
and \emph{robustness} of a practical MLE with multiple quantizers in
the context of \emph{dependent} quantized measurements at the
sensors, unknown parameter \emph{vector} and without the knowledge
of measurement models. The difficulties include the fact that due to
dependence between measurements across sensors, the unknown high
dimensional vector parameter estimation problem cannot be decoupled
to scalar parameter estimation problems; and the quantized samples
are not identically distributed due to the use of multiple different
quantizers. Therefore, we have to deal with unknown \emph{vector}
parameter and unidentically distributed samples simultaneously. The
asymptotic variance is derived to be the inverse of a weighted
linear combination of Fisher information matrices based on $J$
different quantizers which can be used to verify  the robustness of
our approach. A typical estimation problem with a bivariate
non-Gaussian pdf with application to the distributed CFAR detection
systems is considered. Simulations show that the new MLE scheme is
robust and much better than that based on the worst quantization
scheme from among the groups of quantizers. Moreover, when the
number of quantized measurements is small, a surprising result is
that the robust MLE has a significant and dominated advantage over
the MLE with a single quantizer. It is also shown that the
performance of the robust MLE is not the average performance of
multiple quantizers.
The rest of the paper is organized as follows. Problem formulation
is given in Section \ref{sec_2}. In Section \ref{sec_3}, the robust
MLE scheme is proposed and the asymptotic results are derived.
In Section \ref{sec_5},  numerical examples are given and discussed.
In Section \ref{sec_6}, conclusions are made.

\section{Problem formulation}\label{sec_2}
\begin{figure}[h]
\vbox to 5cm{\vfill \hbox to \hsize{\hfill
\scalebox{0.28}[0.3]{\includegraphics{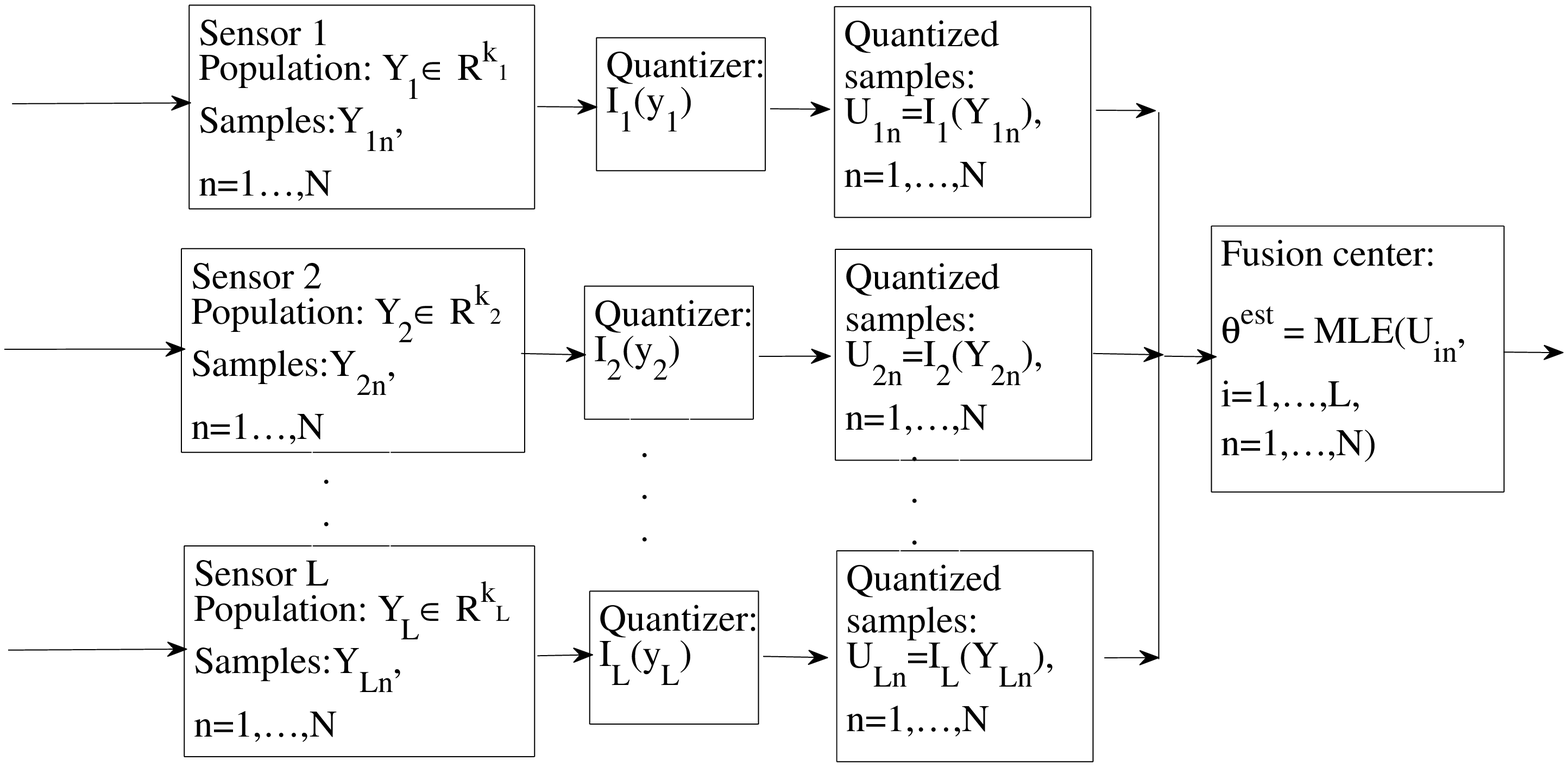}} \hfill}\vfill}
\caption{Distributed MLE fusion system with quantized
data}\label{fig_0}
\end{figure}

The basic $L$-sensor distributed estimation system is considered
(see Figure \ref{fig_0}). Each sensor has  $k_i$-dimensional
observation population $Y_i$, $i=1,\ldots,L$. Suppose that the joint
observation population $Y\triangleq(Y_1',\ldots,Y_L')'$ has a given
family of joint pdf:
\begin{eqnarray}%
\label{Eq2_1} \{p(y_1,\ldots,y_L|\theta)\}_{\theta\in\Theta\subseteq
\mathbb{R}^k}
\end{eqnarray}
where $'$ denotes the transpose and  $\theta$ is the unknown
$k$-dimensional deterministic parameter vector which may include
marginal parameters and dependence parameters. Here, we do not
assume independence across sensors, knowledge of measurement models
and Gaussianity of the joint pdf. Let $N$ independently and
identically distributed (i.i.d.) sensor observation samples and
joint observation samples be
\begin{eqnarray}%
\label{Eq2_2}\vec{Y}_i&=&(Y_{i1},\ldots,Y_{iN}),~ i=1,\ldots,L;\\
\label{Eq2_3} \vec{Y}&=&(\vec{Y}_1',\ldots,\vec{Y}_L')'.
\end{eqnarray}
Suppose the sensors and the fusion center wish to jointly estimate
the unknown parameter vector $\theta$ based on the spatially
distributed observations. If there is sufficient communication
bandwidth and power, the fusion center can obtain  asymptotically
efficient estimates with the complete observation samples based on
MLE procedure under some regularity conditions on the joint pdf.

In many practical situations, however, to reduce the communication
requirement from sensors to the fusion center due to limited
communication bandwidth and power, the $i$-th sensor quantizes the
observation vector to 1 bit (it is straightforward to extend to
multiple bits) by a measurable indicator quantization function:
\begin{eqnarray}%
\label{Eq2_4} && I_i(y_i):~y_i\in\mathbb{R}^{k_i}\rightarrow
\{0,1\},
\end{eqnarray}
for $i=1,\ldots, L$. Here, the quantization region of each quantizer
$I_i(y_i)$ may be  continuous or union of discontinuous regions.
Moreover, we denote by
\begin{eqnarray}%
\label{Eq2_5}\label{Eq2_6}&&I(y)\triangleq
(I_1(y_1),\ldots,I_L(y_L))' \in \mathbb{R}^L.
\end{eqnarray}

Once the  binary quantized samples $I_i(Y_{in})$ are generated at
sensor $i$, $i=1,\ldots,L$, they are transmitted to the fusion
center, for $n=1,\ldots,N$. The fusion center is then required to
estimate the true parameter vector $\theta^*$ based on the received
quantized data.
By the definition of observation samples and quantizers, we define
\begin{eqnarray}%
\label{Eq3_4} \vec{U}&\triangleq&(\vec{U}_{1}',\ldots,\vec{U}_{N}')',\\
\label{Eq3_3} \vec{U}_{n}&\triangleq&(U_{1n},\ldots,U_{Ln})',~
n=1,\ldots,N,\\
\label{Eq3_1} U_{in}&\triangleq& I_i(Y_{in}),~ n=1,\ldots,N.
\end{eqnarray}
If we take $\vec{U}_n$ as the joint quantized observation sample and
denote the quantized observation population by $U\triangleq I(Y)$
$=(I_1(Y_1),$ $\ldots$, $I_L(Y_L))'$, we know that $U$ has a
\emph{discrete/categorical} distribution. Based on the pdf of $Y$
and quantizers $I(y)$, the probability mass function (pmf) of the
quantized observation population $U$ is
\begin{eqnarray}
\nonumber
&&f_U(u_{1},u_{2},\ldots,u_{L}|\theta)\\
\label{Eq3_5}&=&{\int_{\Xi_{(u_{1},u_{2},\ldots,u_{L})}}p(y_{1},y_{2},\ldots,y_{L}|\theta)}dy_{1}dy_{2}\ldots
dy_{L},~~
\end{eqnarray}
where
\begin{eqnarray}
\nonumber(u_{1},u_{2},\ldots,u_{L})&\in& \mathcal {S}_u=\{(u_{1},u_{2},\ldots,u_{L})\in\mathbb{R}^L:\\
\label{Eq3_6}&& u_i=0/1, i=1,\ldots,L\},\\
\nonumber
{\Xi_{(u_{1},u_{2},\ldots,u_{L})}}&=&\{(y_1,y_2,\ldots, y_L):\\
\nonumber&&I_1(y_{1})=u_{1},I_2(y_{2})=u_{2},\\
\label{Eq3_006}&&\ldots, I_L(y_{L})=u_{L}\}.
\end{eqnarray}
Thus, the quantized observation population $U$ has a  family of
joint pmf
$\{f_U(u_{1},u_{2},\ldots,u_{L}|\theta)\}_{\theta\in\Theta\subseteq\mathbb{R}^k}$
which yields the following log likelihood function of samples
$\vec{U}$ by  (\ref{Eq3_4})-(\ref{Eq3_006}):
\begin{eqnarray}%
\label{Eq3_7}l(\theta|\vec{U})
&\triangleq&\log{\prod_{n=1}^Nf_U(U_{1n},U_{2n},\ldots,U_{Ln}|\theta)}\\
\label{Eq3_07} &=&\sum_{n=1}^N\log
f_U(U_{1n},U_{2n},\ldots,U_{Ln}|\theta)\\
\label{Eq3_9} &=&\sum_{j=1}^{2^L}n_j\log{f_U(\vec{u}_j|\theta)}
\end{eqnarray}
where $n_j=\#\{(U_{1n},U_{2n},\ldots,U_{Ln})=\vec{u}_j\in S_u,
n=1,\ldots, N\}$, $\sum_{j=1}^{2^L}n_j=N$; $\#\{\cdot\}$ is the
cardinality of the set. The parameter vector $\theta$ is estimated
by maximizing the log likelihood function (\ref{Eq3_9}). Let
$\hat{\theta}$ denote the MLE of $\theta$.

Based on the classical
asymptotic properties of MLE (see, e.g., textbooks
\cite{Casella-Berger01,VanderVaart00}), we have the following lemma.



\begin{lemma}\label{lem_2}
Assume that $p(y_{1},y_{2},\ldots,y_{L}|\theta)$ and sensor
quantizers $I_1(y_1)$, $\ldots$, $I_L(y_L)$ generate the quantized
samples and $f_U(u_{1},u_{2},\ldots,u_{L}|\theta)$ satisfies the
regularity conditions (A1)--(A6) given on page 516 of
\cite{Casella-Berger01} with respect to the vector parameter
$\theta$; the Fisher information matrix is nonsingular.
%
%
%
Then,
\begin{eqnarray}
\label{Eq3_14} \sqrt{N}(\hat{\theta}-\theta^*)\longrightarrow
\mathcal {N}(0,\mathcal {I}^{-1}(\theta^*,I(y)))
\end{eqnarray}
where $\mathcal {I}^{-1}(\theta^*,I(y))$ is the Cram\'{e}r-Rao lower
bound for one quantized sample which depends on the quantizer
$I(y)$. That is, $\hat{\theta}$ is a consistent and asymptotically
efficient estimator of $\theta^*$.
\end{lemma}

From Lemma \ref{lem_2}, a natural problem that arises is how should
quantizers $I(y)$ be designed such that the asymptotic variance
$\mathcal {I}^{-1}(\theta^*$, $I(y))$ of MLE with quantized data is
as small as possible. The true parameter $\theta^*$, however, is not
known, i.e. the pdf is not known. Most of the existing work on the
design of \emph{optimal} quantizers depends on the availability of
the pdf or signal models. When both of them are not known, an
optimal quantizer cannot be derived or the optimal quantizer depends
on unknown parameters which can not be implemented (see, e.g.,
\cite{Fang-Li08}). Since MLE with a single quantizer is sensitive to
the choice of thresholds due to the uncertainty of pdf, we employ
multiple groups of quantizers (which can be determined by the use of
prior information or some heuristic approach) at each sensor to fend
off against the risk of a single poor/outlier quantizer. To the best
of our knowledge, the asymptotic efficiency and robustness of MLE
scheme with multiple quantizers are not derived analytically in the
context of \emph{dependent} quantized measurements at the  sensors,
unknown parameter \emph{vector} and without the knowledge of
measurement models.


\section{Robust maximum likelihood estimation with quantized
data}\label{sec_3}

The word ``robust" has  many and sometimes inconsistent
connotations. In the theory of robust estimation, robustness
generally means  the ability to resist against outliers  or the
departure from a uncertain model with nominal values and bounds of
uncertainty. In this paper, for our purpose, it means the  ability
to resist against outliers. In this section, we will employ multiple
groups of quantizers  which can be determined by the use of prior
information or some heuristic approach at each sensor to fend off
against the risk of a single poor/outlier quantizer.  The asymptotic
efficiency of the MLE scheme with multiple quantizers is derived
analytically. It enables us to verify and discuss the robustness of
our approach.

The MLE scheme with multiple groups of quantizers  is given as
follows.
\begin{enumerate}
\item Choose $J$ groups of different quantizers
$I^{(j)}(y)\triangleq (I_1^{(j)}(y_1),\ldots,I_L^{(j)}(y_L))' \in
\mathbb{R}^L$, $j=1,\ldots,J,$
where
$I_i^{(j)}(y_i): y_i\in\mathbb{R}^{k_i}\rightarrow \{0,1\}, ~
i=1,\ldots,L.$

\item Observe $N_j$ joint observation samples $\{(Y_{1n_j},\ldots$, $Y_{Ln_j})\}_{n_j=1}^{N_j}$
which are quantized by the $j$-th group of quantizers for
$j=1\ldots,J$. We denote by $N\triangleq \sum_{j=1}^JN_j$. The
quantized observation samples $\{(I_1^{(j)}(Y_{1n_j}),\ldots,$
$I_L^{(j)}(Y_{Ln_j}))\}_{n_j=1}^{N_j}$ are denoted by
$\{(U_{1n_j}^{(j)},$ $\ldots,$ $U_{Ln_j}^{(j)})\}_{n_j=1}^{N_j}$.
Moreover, we denote by $\vec{U}^{(j)}\triangleq\{(U_{1n_j}^{(j)},$
$\ldots,$ $U_{Ln_j}^{(j)})\}_{n_j=1}^{N_j}$. The population of the
quantized sample $(U_{1n_j}^{(j)},$ $\ldots,$ $U_{Ln_j}^{(j)})$ is
denoted by $U^{(j)}$ whose pmf is
\begin{eqnarray}
\label{Eq3_37} f_U^{(j)}(u_{1},u_{2},\ldots,u_{L}|\theta),
j=1,\ldots,J,
\end{eqnarray}
which can be similarly obtained by (\ref{Eq3_5}) and is determined
by $I^{(j)}(y)$ and $p(y_{1},y_{2},\ldots,y_{L}|\theta)$.

\item Estimate the parameter $\theta$ with the $N$ quantized samples which are generated by $J$ groups of
quantizers   by maximizing the log likelihood function:
\begin{eqnarray}%
\nonumber&& l(\theta|\vec{U}^{(1)},\ldots,\vec{U}^{(J)})\\
\label{Eq3_38}&=&\log{\prod_{j=1}^J\prod_{n_j=1}^{N_j}f_U^{(j)}(U_{1n_j}^{(j)},U_{2n_j}^{(j)},\ldots,U_{Ln_j}^{(j)}|\theta)}\\
\label{Eq3_40}&=&\sum_{j=1}^Jl(\theta|\vec{U}^{(j)}),
\end{eqnarray}
where $l(\theta|\vec{U}^{(j)})$ is the log likelihood function of
the $j$-th group of quantized data $\vec{U}^{(j)}$.
Let $\hat{\theta}_{R}$ denote the solution of MLE with $J$
quantizers.
\end{enumerate}

Obviously, the $N$ quantized samples are unidentically distributed
due to the use of $J$ different quantizers.  One may question
whether the new estimator based on the different quantizers is still
asymptotically efficient? What is the asymptotic variance of the new
estimator? Why is it robust compared to using one group of
quantizers? Actually, these questions can be analytically answered
by the following Theorem.

\begin{theorem}\label{thm_2}
There are $J$ groups of different sensor quantizers $I^{(j)}(y)$,
$j=1,\ldots,J$. Assume that $p(y_{1},y_{2},\ldots,y_{L}|\theta)$ and
quantizers $I^{(j)}(y)$ generate the quantized samples and the
quantized  pmf $f_U^{(j)}(u_{1},u_{2},\ldots$, $u_{L}|\theta)$
defined by (\ref{Eq3_37}) satisfies the regularity conditions
(A1)--(A6) given on page 516 of \cite{Casella-Berger01} with respect
to the vector parameter $\theta$ \footnote{Since the   regularity
conditions are fairly standard  and the space is limited, we do not
repeat them in the paper. More discussion on when the regularity
conditions are reasonable can be seen in 10.6.2 of
\cite{Casella-Berger01}.}; the Fisher information matrix is
nonsingular. Then,
\begin{eqnarray}
\label{Eq3_42} \sqrt{N}(\hat{\theta}_{R}-\theta^*)\longrightarrow
\mathcal {N}(0,\mathcal
{I}^{-1}(\theta^*;I^{(1)}(y),\ldots,I^{(J)}(y))
\end{eqnarray}
where $N=\sum_{j=1}^JN_j, N_j\rightarrow\infty$,
$\omega_j=\lim_{N_j\rightarrow\infty}\frac{N_j}{N}, j=1,\ldots,J$,
\begin{eqnarray}
\nonumber&& \mathcal
{I}^{-1}(\theta^*;I^{(1)}(y),\ldots,I^{(J)}(y))\\
\label{Eq3_43}&\triangleq&\left(\sum_{j=1}^J\omega_j\mathcal
{I}(\theta^*;I^{(j)}(y))\right)^{-1},
\end{eqnarray}
$\left(\sum_{j=1}^J\omega_j\mathcal
{I}(\theta^*;I^{(j)}(y))\right)^{-1}$ is the Cram\'{e}r-Rao lower
bound, where $\mathcal {I}(\theta^*;I^{(j)}(y))$ is the Fisher
information matrix for one quantized sample of $U^{(j)}$. That is,
$\hat{\theta}_{R}$ is a consistent and asymptotically efficient
estimator of $\theta^*$.

\end{theorem}

\emph{\textbf{Proof:}}  The regularity of
$p(y_{1},y_{2},\ldots,y_{L}|\theta)$ and quantizers $I^{(j)}(y)$
ensures that the quantized samples and the corresponding pmf
$f_U^{(j)}(u_{1},u_{2},\ldots,u_{L}|\theta)$ defined by
(\ref{Eq3_37}) satisfy the regularity conditions (A1)--(A4) (from
\cite{Casella-Berger01} page 516), and it is easy to prove that
$\hat{\theta}_{R}$   is a consistent estimator of $\theta^*$, i.e.,
$\hat{\theta}_{R}\rightarrow\theta^*,$ in probability. The proof is
similar to that of Theorem 10.1.6 in \cite{Casella-Berger01}.
However, $N$ quantized samples are independent but unidentically
distributed due to the use of $J$ different quantizers. Thus, to
prove the asymptotic normality, we will use the Lyapunov central
limit theorem by checking the Lyapunov condition (see, e.g.,
\cite{VanderVaart00}). Simultaneously, the Cram\'{e}r-Wold device
(see, e.g., \cite{VanderVaart00}) will be used to deal with the high
dimensional estimated parameters.

First, we expand the first derivative of the log likelihood function
(\ref{Eq3_38}) around the true value $\theta^*$,
\begin{eqnarray}
\nonumber &&\frac{\partial
l(\theta|\vec{U}^{(1)},\ldots,\vec{U}^{(J)})}{\partial\theta}\\
\nonumber&=&\left.\frac{\partial
l(\theta|\vec{U}^{(1)},\ldots,\vec{U}^{(J)})}{\partial\theta}\right|_{\theta^*}\\
\nonumber&&+\left.\frac{\partial^2
l(\theta|\vec{U}^{(1)},\ldots,\vec{U}^{(J)})}{\partial\theta^2}\right|_{\theta^*}(\theta-\theta^*)\\
\label{Eq3_47}
&&+\frac{1}{2}D^3(\theta-\theta^*;\theta_0)(\theta-\theta^*),
\end{eqnarray}
where
\begin{eqnarray}
\nonumber&&D^3(\theta-\theta^*;\theta_0)\\
&=&\left(
                                                 \begin{array}{c}
                                                   (\theta-\theta^*)'\{\left.\frac{\partial^2}{\partial\theta^2}\left(\frac{\partial
l(\theta|\vec{U}^{(1)},\ldots,\vec{U}^{(J)})}{\partial\theta_1}\right)\right|_{\theta_0}\} \\
                                                  \vdots \\
                                                   (\theta-\theta^*)'\{\left.\frac{\partial^2}{\partial\theta^2}\left(\frac{\partial
l(\theta|\vec{U}^{(1)},\ldots,\vec{U}^{(J)})}{\partial\theta_k}\right)\right|_{\theta_0}\} \\
                                                 \end{array}
                                               \right)
\end{eqnarray}
$\theta_0$ is between $\theta$ and $\theta^*$.
Substituting $\hat{\theta}_{R}$ for $\theta$ and realizing that the
left-hand of (\ref{Eq3_47}) is \textbf{0} to obtain
\begin{eqnarray}
\nonumber \textbf{0}&=&\left.\frac{\partial
l(\theta|\vec{U}^{(1)},\ldots,\vec{U}^{(J)})}{\partial\theta}\right|_{\hat{\theta}_{R}}\\
\nonumber&=&\left.\frac{\partial
l(\theta|\vec{U}^{(1)},\ldots,\vec{U}^{(J)})}{\partial\theta}\right|_{\theta^*}\\
\nonumber&&+\left.\frac{\partial^2
l(\theta|\vec{U}^{(1)},\ldots,\vec{U}^{(J)})}{\partial\theta^2}\right|_{\theta^*}(\hat{\theta}_{R}-\theta^*)\\
&&+\frac{1}{2}D^3(\hat{\theta}_{R}-\theta^*;\theta_0)(\hat{\theta}_{R}-\theta^*).
\end{eqnarray}
Thus,
\begin{eqnarray}
\nonumber
\sqrt{N}(\hat{\theta}_R-\theta^*)&=&-\left[\frac{1}{N}\left.\frac{\partial^2
l(\theta|\vec{U}^{(1)},\ldots,\vec{U}^{(J)})}{\partial\theta^2}\right|_{\theta^*}\right.\\
\nonumber&&+\left.\frac{1}{2N}D^3(\hat{\theta}_{R}-\theta^*;\theta_0)\right]^{-1}\\
\label{Eq3_48}&&\cdot\frac{1}{\sqrt{N}}\left.\frac{\partial
l(\theta|\vec{U}^{(1)},\ldots,\vec{U}^{(J)})}{\partial\theta}\right|_{\theta^*}.
\end{eqnarray}

Then, we  check  the Lyapunov condition. Denote by
$S_N^2\triangleq\sum_{j=1}^JN_j\mathcal {I}(\theta^*,I^{(j)})$,
$(S_N^\tau)^2\triangleq\sum_{j=1}^JN_j$ $\tau'\mathcal
{I}(\theta^*,I^{(j)})\tau$ for an arbitrary $\tau\neq0$ ($\tau=0$ is
a trivial case), and
\begin{eqnarray}
\label{Eq3_49} \mathcal {M}_j\triangleq
E\left[\left|\tau'\frac{\frac{\partial}
{\partial\theta}{f_U^{(j)}(U_{1n_j},U_{2n_j},\ldots,U_{Ln_j}|\theta)}}{f_U^{(j)}(U_{1n_j},U_{2n_j},\ldots,U_{Ln_j}|\theta)}\right|^3\right]
\end{eqnarray}
which exists, since condition (A3) is satisfied and $U^{(j)}$ is a
categorical distribution. Moreover, by  condition (A5) and
(\ref{Eq3_49}),
\begin{eqnarray}
\nonumber
&&\lim_{N\rightarrow\infty}\frac{1}{(S_N^{\tau})^3}\sum_{j=1}^J\sum_{n_j=1}^{N_j}\\
\nonumber&&E\left[\left|\tau'\frac{\partial}
{\partial\theta}\log{f_U^{(j)}(U_{1n_j},U_{2n_j},\ldots,U_{Ln_j}|\theta)}\right.\right.\\
\nonumber&&-\left.\left. E[\tau'\frac{\partial}
{\partial\theta}\log{f_U^{(j)}(U_{1n_j},U_{2n_j},\ldots,U_{Ln_j}|\theta)}]\right|^3\right]\\
\nonumber&=&\lim_{N\rightarrow\infty}\frac{1}{(S_N^{\tau})^3}\sum_{j=1}^J\sum_{n_j=1}^{N_j}\\
\nonumber&&E\left[\left|\tau'\frac{\partial}
{\partial\theta}\log{f_U^{(j)}(U_{1n_j},U_{2n_j},\ldots,U_{Ln_j}|\theta)}-0\right|^3\right]\\
\nonumber&\leq&\lim_{N\rightarrow\infty}\frac{1}{(S_N^{\tau})^3}\sum_{j=1}^J\sum_{n_j=1}^{N_j}\mathcal
{M}_j\\
\nonumber&\leq&\lim_{N\rightarrow\infty}\frac{1}{(S_N^{\tau})^3}N\max\{\mathcal
{M}_1,\ldots,\mathcal {M}_J\}\\
\nonumber&\leq&\lim_{N\rightarrow\infty}\frac{N\max\{\mathcal
{M}_1,\ldots,\mathcal {M}_J\}}{(N\min\{\tau'\mathcal
{I}(\theta^*,I^{(1)})\tau,\ldots,\tau'\mathcal
{I}(\theta^*,I^{(J)})\tau\})^{\frac{3}{2}}}\\
\nonumber&\leq&\lim_{N\rightarrow\infty}\frac{1}{\sqrt{N}}\frac{\max\{\mathcal
{M}_1,\ldots,\mathcal {M}_J\}}{\min\{\tau'\mathcal
{I}(\theta^*,I^{(1)})\tau,\ldots,\tau'\mathcal
{I}(\theta^*,I^{(J)})\tau\}}\\
\nonumber&=&0.
\end{eqnarray}
That is, the Lyapunov condition  is satisfied. Thus, by the Lyapunov
central limit theorem (see, e.g., \cite{VanderVaart00}), for all
$\tau$,
\begin{eqnarray}
\nonumber \frac{1}{\sqrt{N}}\tau'\left.\frac{\partial
l(\theta|\vec{U}^{(1)},\ldots,\vec{U}^{(J)})}{\partial\theta}\right|_{\theta^*}\rightarrow
\mathcal {N}(0, (S_\omega^{\tau})^2)~~~\\
 (\mbox{in
distribution}),\nonumber
\end{eqnarray}
where $(S_\omega^{\tau})^2 \triangleq\sum_{j=1}^J\omega_j
\tau'\mathcal {I}(\theta^*,I^{(j)})\tau$.  Moreover, by the
Cram\'{e}r-Wold device (see, e.g., \cite{VanderVaart00}), we have
\begin{eqnarray}
\label{Eq3_50}  \frac{1}{\sqrt{N}}\left.\frac{\partial
l(\theta|\vec{U}^{(1)},\ldots,\vec{U}^{(J)})}{\partial\theta}\right|_{\theta^*}\rightarrow
\mathcal {N}(0, S_\omega^2) ~~~ \\(\mbox{in distribution})\nonumber,
\end{eqnarray}
where $S_\omega^2 \triangleq\sum_{j=1}^J\omega_j \mathcal
{I}(\theta^*,I^{(j)})$. By application of the weak law of large
number, we have
\begin{eqnarray}
\label{Eq3_51}  \frac{1}{N_j}\left.\frac{\partial^2
l(\theta|\vec{U}^{(j)})}{\partial\theta^2}\right|_{\theta^*}\rightarrow
-\mathcal {I}(\theta^*;I^{(j)}(y)), j=1,\ldots,J,~~~
\\(\mbox{in probability})\nonumber
\end{eqnarray}
where $l(\theta|\vec{U}^{(j)})$ is defined in (\ref{Eq3_40}). By
Slutsky's Theorem and Equation (\ref{Eq3_51}), we have
\begin{eqnarray}
\nonumber&& \frac{1}{N}\left.\frac{\partial^2
l(\theta|\vec{U}^{(1)},\ldots,\vec{U}^{(J)})}{\partial\theta^2}\right|_{\theta^*}\\
\nonumber&=&\sum_{j=1}^J\frac{N_j}{N}\frac{1}{N_j}\left.\frac{\partial^2
l(\theta|\vec{U}^{(j)})}{\partial\theta^2}\right|_{\theta^*}\\
\label{Eq3_52}&\rightarrow& -\sum_{j=1}^J\omega_j\mathcal
{I}(\theta^*;I^{(j)}(y))=-S_\omega^2~ (\mbox{in probability}).
\end{eqnarray}
Since condition (A6) given on page 516 of \cite{Casella-Berger01}
guarantees that three times differentiation of the log likelihood
function can be bounded by an integrable function for all $\theta$
in a small neighborhood of $\theta^*$ and note that $\theta_0$ is
between $\hat{\theta}_{R}$ and $\theta^*$,
$\hat{\theta}_{R}\rightarrow\theta^*$ (in probability), we have
\begin{eqnarray}
\label{Eq3_052}
\frac{1}{2N}D^3(\hat{\theta}_{R}-\theta^*;\theta_0)\rightarrow
\textbf{0}~~~(\mbox{in probability})
\end{eqnarray}
Moreover, based on Equations (\ref{Eq3_48}) (\ref{Eq3_50}),
(\ref{Eq3_52}), (\ref{Eq3_052}) and Slutsky's Theorem, we have
\begin{eqnarray}
\label{Eq3_53} \sqrt{N}(\hat{\theta}_{R}-\theta^*)\longrightarrow
\mathcal {N}(0,\mathcal
{I}^{-1}(\theta^*;I^{(1)}(y),\ldots,I^{(J)}(y)) ~~~ \\
(\mbox{in distribution})\nonumber
\end{eqnarray}
where $\mathcal {I}^{-1}(\theta^*;I^{(1)}(y),\ldots,I^{(J)}(y))$
defined by (\ref{Eq3_43}) and is the Cram\'{e}r-Rao lower bound.
Therefore, $\hat{\theta}_{R}$ is a consistent and asymptotically
efficient estimator of $\theta^*$. ~$\Box$

\begin{remark}\label{rmk_02}

As we have shown that the asymptotic variance of multiple quantizers
is the inverse of a weighted mean of Fisher information matrices
based on $J$ different quantizers. Without loss of generality,
assume that the weights are equal and the first quantizer is an
outlier, i.e., the asymptotic variance of multiple quantizers   is
the inverse of the mean  of Fisher information matrices based on $J$
different quantizers and the asymptotic variance
${I}(\theta^*;I^{(1)}(y))^{-1}$ is \emph{much larger} than the other
$J-1$ asymptotic variances ${I}(\theta^*;I^{(j)}(y))^{-1},
j=2,\ldots,J$. Since ${I}(\theta^*;I^{(j)}(y))^{-1}, j=1,\ldots,J$
are positive definite matrices and $A\succeq B$ implies
$A^{-1}\preceq B^{-1}$ for positive definite matrices, the Fisher
information  ${I}(\theta^*;I^{(1)}(y)) $ is \emph{much smaller} than
the other $J-1$ Fisher informations ${I}(\theta^*;I^{(j)}(y)),
j=2,\ldots,J$ respectively. Thus, the mean  of Fisher informations
with outlier $\frac{1}{J}\sum_{j=1}^J{I}(\theta^*;I^{(j)}(y))$ and
that without outlier
$\frac{1}{J-1}\sum_{j=2}^J{I}(\theta^*;I^{(j)}(y))$ are \emph{much
larger} than ${I}(\theta^*;I^{(1)}(y))$ and are very close to each
other with the same order of magnitude. Moreover,  the corresponding
asymptotic variances are \emph{much smaller} than
${I}(\theta^*;I^{(1)}(y))^{-1}$ and are very close to each other
with the same order of magnitude by the continuity of matrix
inverse. Therefore, the MLE scheme with multiple quantizers is a
robust scheme.

As a simple numerical  example, let us consider that there are 3
quantizers with asymptotic variances  $\frac{1}{3}\times 10^3$,
$\frac{1}{3}, \frac{1}{3.3}$ respectively. Obviously, the first
quantizer is an outlier. It can be calculated that the asymptotic
variance of robust MLE when equally using 3 different quantizers is
 $\frac{1}{\frac{1}{3}(3 \times 10^{-3}+3+3.3)}= 0.4760$ which is much smaller than
that of the outlier $\frac{1}{3}\times 10^3$ and has the same order
of magnitude as $\frac{1}{ \frac{1}{2}(3+3.3)}= 0.3175$ and
$\frac{1}{3}, \frac{1}{3.3}$.
\end{remark}

\section{Numerical Examples}\label{sec_5}
In distributed detection systems, the detection performance relies
heavily on the knowledge of the joint pdf under hypotheses $H_0$ and
$H_1$. Here, we consider the problem of estimating joint pdf under
$H_1$ for distributed CFAR  detection systems \cite{Varshney97} that
has great practical relevance. In these systems, the marginal
distribution of measurements is usually assumed exponentially
distributed or Gamma pdf. By noting that the exponential pdf is a
special case of the Gamma pdf, we consider the marginals of a
two-sensor system to follow a Gamma distribution as follows:
\begin{eqnarray}
\nonumber S_i:&& Y_i\sim Gamma(\theta_i,4),~\\
\nonumber&&
p_i(y_i|\theta_i)=\frac{y_i^{\theta_i-1}e^{-y_i/4}}{4^{\theta_i}\Gamma(\theta_i)},
\theta_i>0, i=1,2,
\end{eqnarray}
where  $ \theta_1$ and $\theta_2$ are the parameters to be
estimated. It has been shown recently that the dependence between
sensors is very important to the distributed detection performance
(see e.g., \cite{Iyengar-Varshney-Damarla11}). To estimate the
dependence between sensors,  copula theory can be used to construct
the structure of dependence. By Sklar's Theorem in copula therory
(see, e.g., \cite{Nelsen99}), the joint pdfs can be written as
follows:
\begin{eqnarray}
\nonumber
&&p(y_1,y_2|\theta)=c(F_1(y_1|\theta_1),F_2(y_2|\theta_2))|\theta_0)\prod_{i=1}^2p_i(y_i|\theta_i),
\end{eqnarray}
where $p_i(y_i|\theta_i)$ and $F_i(y_i\theta_i)$ are marginal pdf
and cumulative distribution function respectively; $c(v_1,v_2$
$|\theta_0)$ is the copula density. For a specific numerical
example, we consider
the joint Clayton copula density  as follows:
\begin{eqnarray}
\nonumber&& c(v_1,v_2|\theta_0)\\
\nonumber&=&(1+\theta_0)v_1^{-1-\theta_0}v_2^{-1-\theta_0}\\
\nonumber&&
(-1+v_1^{-\theta_0}+v_2^{-\theta_0})^{-2-1/\theta_0},\theta_0\in[-1,\infty)\backslash\{0\},
\end{eqnarray}
which is a frequently used copula model to describe dependence (see
\cite{Nelsen99}). The parameter vector to be estimated is
$\theta\triangleq[\theta_0, \theta_1, \theta_2]$ corresponding to
the copula density and the two marginals.
We compare the robust MLE with MLE based on a single quantizer. We
assume that the prior information is that the thresholds are in
$[10, 25]$. Based on this information, we uniformly choose the
following four groups of different quantizers.
\begin{eqnarray}
\nonumber I^{(1)}(y)=(I_1^{(1)}(y_1), I_2^{(1)}(y_2))=(I[y_1-25],
I[y_2-25]),\\
\nonumber I^{(2)}(y)=(I_1^{(2)}(y_1), I_2^{(2)}(y_2))=(I[y_1-20],
I[y_2-20]),\\
\nonumber I^{(3)}(y)=(I_1^{(3)}(y_1), I_2^{(3)}(y_2))=(I[y_1-15],
I[y_2-15]),\\
\nonumber I^{(4)}(y)=(I_1^{(4)}(y_1), I_2^{(4)}(y_2))=(I[y_1-10],
I[y_2-10]),
\end{eqnarray}
where  $I[x-c]=1$ if $x\geq c$; otherwise
$I[x-c]=0$.
For the robust MLE, we let $N_1=N_2=N_3=N_4=N/4$ where $N$ is the
number of samples for MLE with fixed quantizer $I^{(j)}(y),
j=1,2,3,4$ respectively.

\begin{figure}[h]
\vbox to 5cm{\vfill \hbox to \hsize{\hfill
\scalebox{0.275}[0.275]{\includegraphics{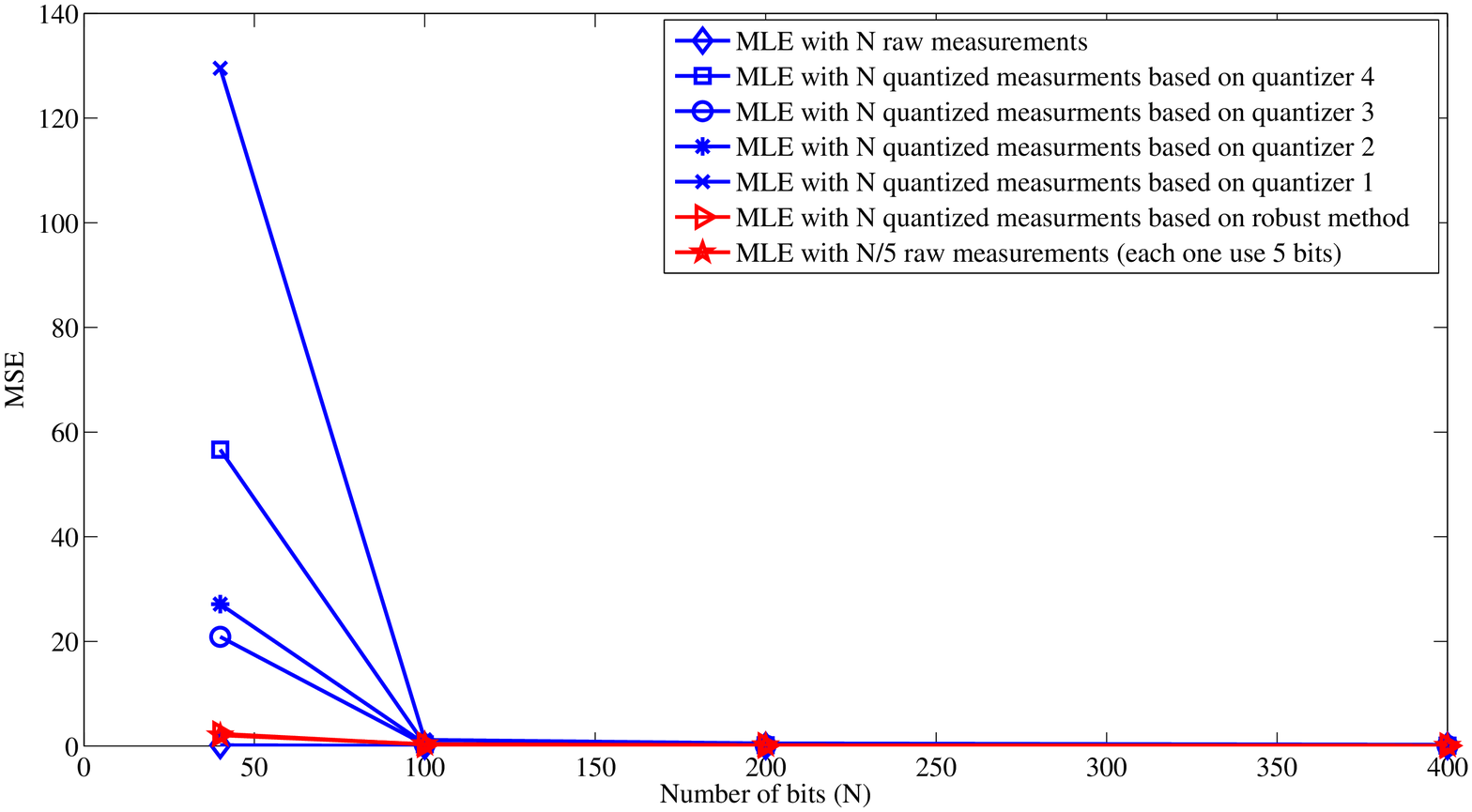}} \hfill}\vfill}
\caption{MSEs of MLE of $\theta_0$ based on 1000 M.C. runs while
using raw measurements, different quantizers and the robust MLE of
$\theta_0$ for different number of measurements. Figure \ref{fig_4}
is using linear scale.}\label{fig_4}
\end{figure}

\begin{figure}[h]
\vbox to 5cm{\vfill \hbox to \hsize{\hfill
\scalebox{0.275}[0.275]{\includegraphics{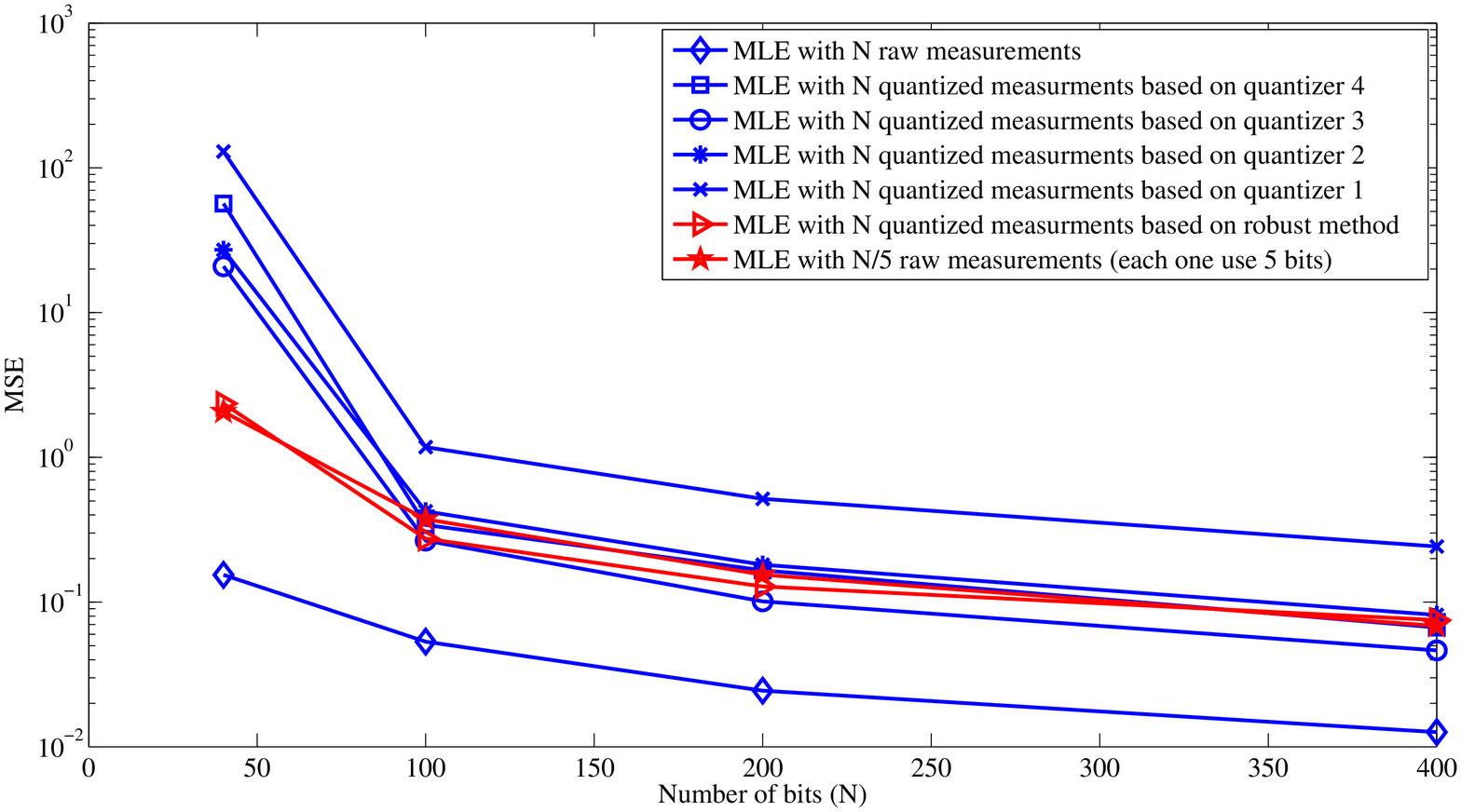}} \hfill}\vfill}
\caption{Figure \ref{fig_4} using logarithmic scale.}\label{fig_04}
\end{figure}

\begin{figure}[h]
\vbox to 5cm{\vfill \hbox to \hsize{\hfill
\scalebox{0.275}[0.275]{\includegraphics{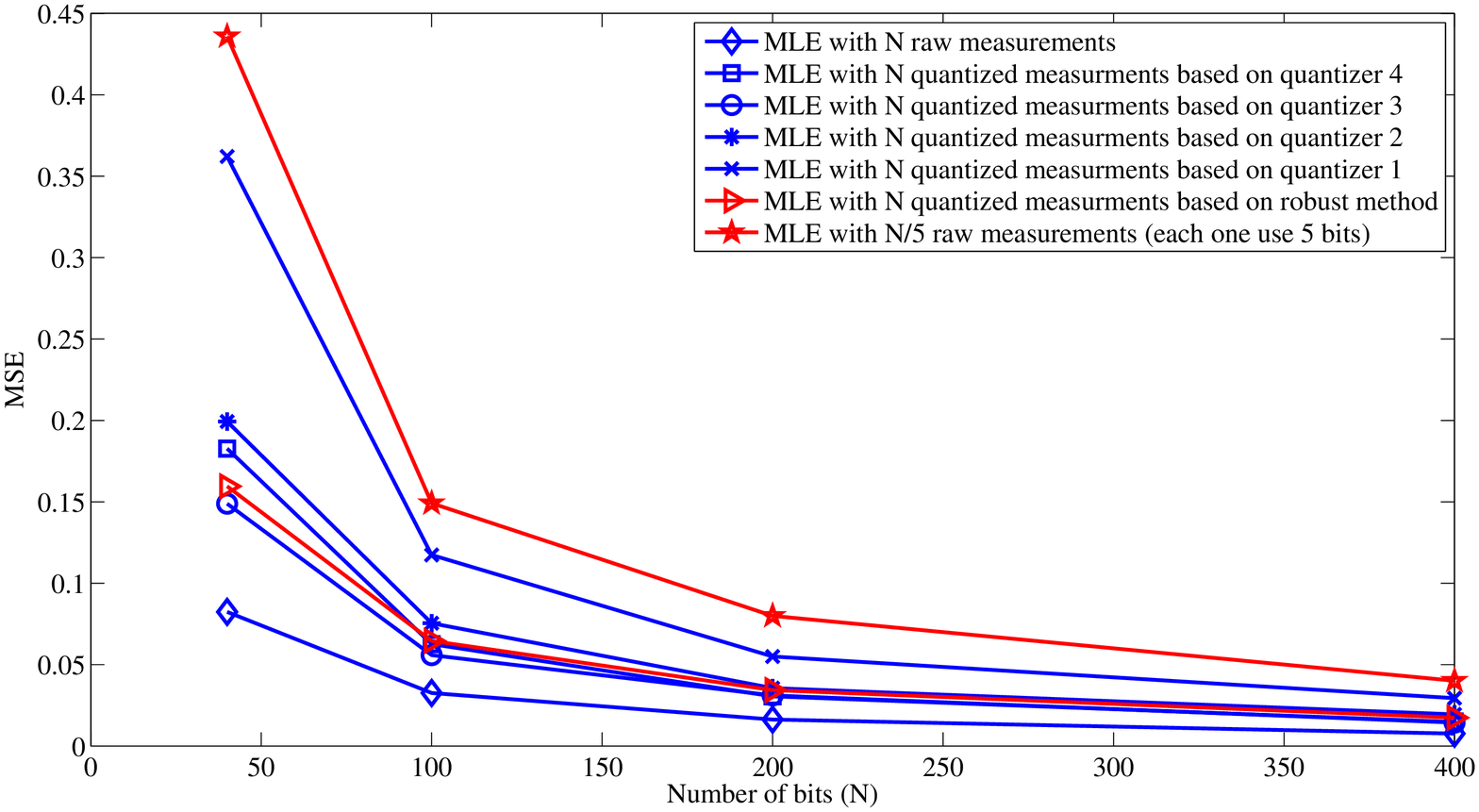}} \hfill}\vfill}
\caption{MSEs of MLE of $\theta_1$ based on 1000 M.C. runs while
using raw measurements, different quantizers and the robust MLE of
$\theta_1$ for different number of measurements. }\label{fig_5}
\end{figure}

\begin{figure}[h]
\vbox to 5cm{\vfill \hbox to \hsize{\hfill
\scalebox{0.275}[0.275]{\includegraphics{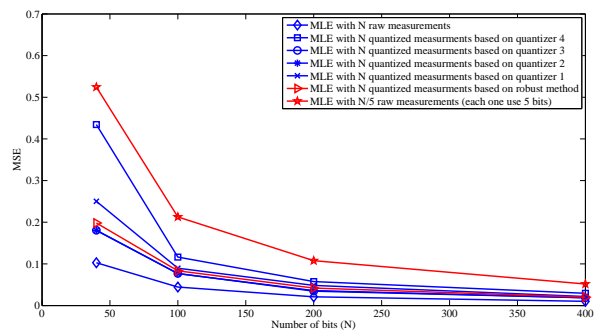}} \hfill}\vfill}
\caption{MSEs of MLE of $\theta_2$ based on 1000 M.C. runs while
using raw measurements, different quantizers and the robust MLE of
$\theta_2$ for different number of measurements.}\label{fig_6}
\end{figure}

The robustness of the MLE with multiple quantizers is illustrated in
Figs. \ref{fig_4}--\ref{fig_6}.
In Figs. \ref{fig_4}--\ref{fig_6}, MSEs based on 1000 Monte Carlo
(M.C.) runs as a function of the number of  measurements $N=[40,
100, 200, 400]$  for different estimation methods (MLE with single
quantizer, robust MLE and MLE with raw measurements) are plotted for
parameters $\theta_0=1.0759$, $\theta_1=4$ and $\theta_2=5$
respectively, where $\theta_0=1.0759$ corresponds to the dependence
measure namely Spearman's $\rho=0.5$.
Figs \ref{fig_4}--\ref{fig_04} present the   MSEs of $\theta_0$ in
linear and logarithmic scales respectively. In Figs
\ref{fig_5}--\ref{fig_6}, we present the MSEs of $\theta_1$ and
$\theta_2$ in linear scale respectively. In our work thus far, we
have assumed that 1-bit quantized data is transmitted to the fusion
center in simulations. We consider another system where finely
quantized data (5-bit data corresponding to a subset of samples
instead of 1-bit data corresponding to all the samples) is
transmitted while maintaining the total number of bits equal to $N$.
While evaluating the performance of the system with 5-bit data, we
employ the results corresponding to raw data which, in fact, give
more optimistic results. The MSEs based on 1000 M.C. runs of
transmitting $N/5$ finely quantized 5-bit measurements are given in
Figs \ref{fig_4}--\ref{fig_6} for $\theta_0$, $\theta_1$ and
$\theta_2$ respectively.



From Figs \ref{fig_4}--\ref{fig_6}, we have the following
observations:
(1). From Figs \ref{fig_4}--\ref{fig_6},  MSEs based on 1000 M.C.
runs for robust MLE are much smaller than those of the MLE based on
the single quantizer  that is the worst (outlier) in the group. This
phenomenon is consistent with the results in Theorem \ref{thm_2} and
Remark \ref{rmk_02}. Robust MLE is a  conservative estimate, but it
can avoid large errors in the worst case. The advantage of
robustness (MSE of the worst MLE minus MSE of Robust MLE) is much
larger than the loss due to conservative estimation to enhance
robustness (MSE of Robust MLE minus MSE of the best one), especially
in Figs   \ref{fig_4}, \ref{fig_04} and \ref{fig_5}. (2). From Figs
\ref{fig_4}--\ref{fig_04}, a surprising result that is observed is
that the Robust MLE based on 1000 M.C. runs has a significant
advantage over MLE with a single quantizer, when the number of
quantized measurements is small N=40. The reason is that, for small
number of samples, the MLE with single quantizer is sensitive to the
randomized samples so that it may be an outlier in each M.C. run
resulting in poor performance.
(3). By comparing our robust MLE with 1-bit quantized data with MLE
that transmits a subset of finely quantized data in Figs
\ref{fig_4}--\ref{fig_04}, we observe that their performance of
estimating $\theta_0$ is very close. However, for the performance of
estimating $\theta_1$ and $\theta_2$, robust MLE is much better than
the latter from   Figures \ref{fig_5}--\ref{fig_6}. Thus, robust MLE
is a better estimation method in distributed dynamic systems with
limited bandwidth.


\section{Conclusion}\label{sec_6}
In this paper, we have proposed an approach for robust distributed
MLE with dependent quantized data  under the assumption that the
structure of the joint pdf is known, but it contains unknown
deterministic parameters.
We considered a practical estimation problem with a bivariate
non-Gaussian pdf arising from the distributed constant false alarm
rate (CFAR) detection systems. Simulation results show that the new
MLE scheme is robust and much better than that based on the worst
(outlier) quantization scheme from among the groups of quantizers.
An important obersvation is that the robust MLE has a significant
advantage over MLE with a single quantizer, when the number of
quantized measurements is small.


\section*{Acknowledgment}
We would like to thank the anonymous reviewers, the associate editor
and Dr. Kush Varshney of IBM for their helpful suggestions that
greatly improved the quality of this paper.

\begin{thebibliography}{10}

\bibitem{Casella-Berger01}
George Casella and Roger~L. Berger.
\newblock {\em Statistical Inference}.
\newblock Duxbury, New York, second edition, 2001.

\bibitem{Casini-Garulli-Vicino12}
Marco Casini, Andrea Garulli, and Antonio Vicino.
\newblock Input design in worst-case system identification with quantized
  measurements.
\newblock {\em Automatica}, 48:2997--3007, 2012.

\bibitem{Fang-Li08}
Jun Fang and Hongbin Li.
\newblock Distributed adaptive quantization for wireless sensor networks: From
  delta modulation to maximum likelihood.
\newblock {\em IEEE Transactions on Signal Processing}, 56(10):5246--5257,
  2008.

\bibitem{Fang-Li09}
Jun Fang and Hongbin Li.
\newblock Hyperplane-based vector quantization for distributed estimation in
  wireless sensor networks.
\newblock {\em IEEE Transactions on Information Theory}, 55:5682--5699, 2009.

\bibitem{Godoy-Goodwin-Aguero-Marelli-Wigren11}
Boris~I. Godoy, Graham~C. Goodwin, Juan~C. Aguero, Damian Marelli,
and Torbjorn
  Wigren.
\newblock On identification of fir systems having quantized output data.
\newblock {\em Automatica}, 47:1905--1915, 2011.

\bibitem{Gustafsson-Karlsson09}
Fredrik Gustafsson and Rickard Karlsson.
\newblock Statistical results for system identification based on quantized
  observations.
\newblock {\em Automatica}, 45:2794--2801, 2009.

\bibitem{Iyengar-Varshney-Damarla11}
Satish~G. Iyengar, Pramod~K. Varshney, and Thyagaraju Damarla.
\newblock A parametric copula-based framework for hypothesis testing using
  heterogeneous data.
\newblock {\em IEEE Transactions on Signal Processing}, 59(5):2308--2319, May
  2011.

\bibitem{Marelli-You-Fu13}
Dami¨¢n Marelli, Keyou You, and Minyue Fu.
\newblock Identification of {ARMA} models using intermittent and quantized
  output observations.
\newblock {\em Automatica}, 49:360--369, 2013.

\bibitem{Megalooikonomou-Yesha00}
V.~Megalooikonomou and Y.~Yesha.
\newblock Quantizer design for distributed estimation with communication
  constraints and unknown observation statistics.
\newblock {\em IEEE Transactions on Communications}, 48(2):181--184, February
  2000.

\bibitem{Nelsen99}
R.~B. Nelsen.
\newblock {\em An Introduction to Copulas}.
\newblock Springer-Verlag, New York, 1999.

\bibitem{Ribeiro-Giannakis05}
A.~Ribeiro and G.~B. Giannakis.
\newblock Bandwidth-constrained distributed estimation for wireless sensor
  networks--{Part I}: Gaussian case.
\newblock {\em IEEE Transactions on Signal Processing}, 54(3):1131--1143, March
  2005.

\bibitem{Spall12}
James~C. Spall.
\newblock Asymptotic normality and uncertainty bounds for reliability estimates
  from subsystem and full system tests.
\newblock In {\em American Control Conference}, pages 56--61, Fairmont Queen
  Elizabeth, Montreal, Canada, June 2012.

\bibitem{Sundaresan-Varshney11}
Ashok Sundaresan and Pramod~K. Varshney.
\newblock Location estimation of a random signal source based on correlated
  sensor observations.
\newblock {\em IEEE Transactions on Signal Processing}, 59(2):787--799,
  February 2011.

\bibitem{VanderVaart00}
A.~W. {Van der Vaart}.
\newblock {\em Asymptotic statistics}.
\newblock Cambridge University Press, New York, 2000.

\bibitem{Varshney97}
Pramod~K. Varshney.
\newblock {\em Distributed Detection and Data Fusion}.
\newblock New York: Springer-Verlag, 1997.

\bibitem{Veeravalli-Varshney12}
Venugopal~V. Veeravalli and Pramod~K. Varshney.
\newblock Distributed inference in wireless sensor networks.
\newblock {\em Philosophical Transactions of the Royal Society A: Mathematical,
  Physical and Engineering Sciences}, 370(1958):100--117, January 2012.

\bibitem{Wang-Yin-Zhang-Zhao10}
Le~Yi Wang, G.~George Yin, Ji-Feng Zhang, and Yanlong Zhao.
\newblock {\em System Identification with Quantized Observations}.
\newblock Birkhauser, 2010.

\bibitem{Wigren95}
Torbjorn Wigren.
\newblock Approximate gradients, convergence and positive realness in recursive
  identification of a class of non-linear systems.
\newblock {\em International Journal of Adaptive Control and Signal
  Processing}, 9:325--354, 1995.

\bibitem{Xiao-Ribeiro-Luo-Giannakis06}
Jin-Jun Xiao, Alejandro Ribeiro, Zhi-Quan Luo, and Georgios~B.
Giannakis.
\newblock Distributed compression-estimation using wireless sensor networks.
\newblock {\em IEEE Signal Processing Magazine}, 23(4):27--41, July 2006.

\end{thebibliography}
\end{document}